\documentclass[10pt, conference, compsocconf]{IEEEtran}
\usepackage{cite}

\ifCLASSINFOpdf
\else
\fi

\usepackage{graphicx}

\begin{document}
%
\title{Novel and topical business news and their impact on stock market activities}




%
\author{\IEEEauthorblockN{Takayuki Mizuno\IEEEauthorrefmark{1}\IEEEauthorrefmark{4}\IEEEauthorrefmark{5}\IEEEauthorrefmark{6},
Takaaki Ohnishi\IEEEauthorrefmark{2}\IEEEauthorrefmark{6},
Tsutomu Watanabe\IEEEauthorrefmark{3}\IEEEauthorrefmark{6}}
\IEEEauthorblockA{\IEEEauthorrefmark{1}National Institute of Informatics, Tokyo, Japan\\
mizuno@nii.ac.jp}
\IEEEauthorblockA{\IEEEauthorrefmark{2}Graduate School of Information Science and Technology, University of Tokyo, Tokyo, Japan\\
ohnishi.takaaki@i.u-tokyo.ac.jp}
\IEEEauthorblockA{\IEEEauthorrefmark{3}Graduate School of Economics, University of Tokyo, Tokyo, Japan\\
watanabe@e.u-tokyo.ac.jp}
\IEEEauthorblockA{\IEEEauthorrefmark{4}Department of Informatics, SOKENDAI (The Graduate University for Advanced Studies), Tokyo, Japan}
\IEEEauthorblockA{\IEEEauthorrefmark{5}PRESTO, Japan Science and Technology Agency, Tokyo, Japan}
\IEEEauthorblockA{\IEEEauthorrefmark{6}The Canon Institute for Global Studies, Tokyo, Japan}}


\maketitle

\begin{abstract}
We propose an indicator to measure the degree to which a particular news article is novel, as well as an indicator to measure the degree to which a particular news item attracts attention from investors. The novelty measure is obtained by comparing the extent to which a particular news article is similar to earlier news articles, and an article is regarded as novel if there was no similar article before it. On the other hand, we say a news item receives a lot of attention and thus is highly topical if it is simultaneously reported by many news agencies and read by many investors who receive news from those agencies. The topicality measure for a news item is obtained by counting the number of news articles whose content is similar to an original news article but which are delivered by other news agencies. To check the performance of the indicators, we empirically examine how these indicators are correlated with intraday financial market indicators such as the number of transactions and price volatility. Specifically, we use a dataset consisting of over 90 million business news articles reported in English and a dataset consisting of minute-by-minute stock prices on the New York Stock Exchange and the NASDAQ Stock Market from 2003 to 2014, and show that stock prices and transaction volumes exhibited a significant response to a news article when it is novel and topical. 

\end{abstract}


%
\IEEEpeerreviewmaketitle

\section{Introduction}
Financial markets can be regarded as a non-equilibrium open system. Understanding how they work remains a great challenge to researchers in finance, economics, and statistical physics. Fluctuations in financial market prices are sometimes driven by endogenous forces and sometimes by exogenous forces. Business news is a typical example of exogenous forces. Casual observation indicates that stock prices respond to news articles reporting on new developments concerning companies' circumstances. Market reactions to news have been extensively studied by researchers in several different fields \cite{Ito, Chan, Vega, DellaVigna, Petersen_1, Petersen_2, Mitra, Rangel, Engelberg, Mizuno, Elder, Smales, Storkenmaier}, with some researchers attempting to construct models that capture static and/or dynamic responses to endogenous and exogenous shocks \cite{Filimonov_1, Filimonov_2}. The starting point for neoclassical financial economists typically is what they refer to as the ``efficient market hypothesis,'' which implies that stock prices respond at the very moment that news is delivered to market participants. A number of empirical studies have attempted to identify such an immediate price response to news but have found little evidence supporting the efficient market hypothesis \cite{Cutler, McQueen, Fleming, Fair, Joulin, Erdogan}.

Investors seek to forecast what will happen in the near future, and buy and sell securities based on such forecasts. Through this process, some newsworthy developments are factored into market prices before they occur, so that stock prices do not respond at all when they are reported \cite{Bomfim}. This means that it is important for researchers to distinguish between anticipated and unanticipated news and focus only on unanticipated news in detecting the immediate response to news. To do this, we need to measure the extent to which a news article is novel to market participants, which is the first issue we will discuss in this paper. On the other hand, even if a particular piece of news is unanticipated, market responses differ depending on the importance of that piece of news to market participants. Specifically, it has been shown that market reaction to news differs depending on how it is interpreted by market participants \cite{Birz}, on how it is reported by the media (i.e., whether it is reported in a pessimistic or an optimistic context) \cite{Tetlock}, and on how many times the same news item is reported \cite{Lillo}. It has also been shown that transaction volumes tend to be greater for stocks with a larger number of searches on the internet \cite{Bordino}. All of these pieces of evidence suggest that we need to distinguish news that attract a lot of attention from market participants and news that receive little attention, and focus on news attracting a lot of attention in assessing the market response to such news. This means that we need to measure the extent to which a news item attracts attention from market participants, which is the second issue we will discuss in this paper.

Our approach to measure the novelty and topicality of news is closely related to recent studies on the application of text mining techniques to the analysis of financial market activities. Specifically, it has been shown that linguistic and statistical characteristics of news articles extracted using text mining techniques contain useful information to predict future stock prices and trading volumes \cite{Schumaker, Bollen, Mian, Hisano, Ranco, Luss}. Also, in the context of information filtering, several new methods for detecting and eliminating redundant text in blogs and on twitter have been developed and applied to the identification of the novelty content of social networking service (SNS) texts \cite{Zhang, Gabrilovich, Zhao, Ng, Liang}. Our paper is most closely related to studies by the Thomson Reuters Corporation, which propose to measure the novelty of news by counting the number of linguistically similar news articles that are found in a particular time period in their news products \cite{Reuters_1, Reuters_2}. Based on this method, it was shown that financial market activities respond more strongly to follow-up news than to initial news \cite{Gross-Klussmann}. Another study closely related to ours is ref~\cite{Phuvipadawat}, which attempts to measure the importance of a news article by counting the number of retweets of a tweet mentioning the article \cite{Phuvipadawat}.

In this paper, we measure the novelty of a news article by comparing it with other news articles reported before that article in terms of linguistic similarity: the article is regarded as novel if there was no linguistically similar news article before it. This approach is almost the same as that taken in previous studies. On the other hand, we say a news item attracts a lot of attention and thus is highly topical if it is simultaneously reported by multiple news agencies and read by many investors who acquire news from those agencies. The topicality measure for a news article is obtained by counting the number of news articles which have a similar content to the original news article but are delivered by other news agencies. This measure is similar to the measure proposed by ref~\cite{Phuvipadawat} but differs from it in that our measure is able to capture the extent to which a news article is topical immediately after it is delivered, while the measure proposed by ref~\cite{Phuvipadawat} does not work that quickly because the number of retweets of a tweet mentioning the article increases only gradually. To check the performance of the indicators, we empirically examine how they are correlated with intraday financial market indicators such as the number of transactions and price volatility. Specifically, we use a dataset consisting of over 90 million business news articles reported in English and a dataset consisting of minute-by-minute stock prices on the New York Stock Exchange (NYSE) and the NASDAQ Stock Market from 2003 to 2014, and show that stock prices and transaction volumes exhibited a significant response to a news article when it is novel and topical.

The rest of the paper is organized as follows. We first provide a detailed description of our dataset containing over 90 million English-language business news articles, and show that breaking news have much more impact on stock prices and transaction volume than other news. Next, we examine the statistical laws regarding linguistic similarity among news articles, and propose a measure for the novelty of a news article as well as a measure for the topicality of an article. We then examine how these indicators are correlated with intraday financial market indicators.

\section{News dataset}
The Thomson Reuters Corporation (RTRS) and the Dow Jones \& Company Inc. (DJ) deliver news to market participants around the world within fractions of a second through electronic systems \cite{Reuters_3, Dow}. News items published by over 300 third parties are displayed on RTRS's electronic trading platform. In this paper, we use only English-language news articles published by RTRS, the Business Wire News Service (BSW), the Canada Newswire News Service (CNW), Marketwire (MKW), the PR Newswire News Service (PRN), and Market News Publishing Inc. (VMN) on RTRS's platform as well as all of the English-language news articles by DJ from 2003 to 2014. The total number of news articles exceeds 90 million. Journalists include keywords in their articles on RTRS's platform. For example, news articles for General Motors Company, LLC have a keyword, GM.N, where .N means the New York stock exchange (NYSE). There are three types of news events on RTRS's platform. ALERT articles, which provide a one-line summary of breaking news, are displayed in red. HEADLINE articles provide a one-line summary of non-breaking news. An ALERT and a HEADLINE are up to 80-100 characters long. A STORY shows a complete news article. The percentages of ALERTs and HEADLINEs in our dataset are about 12\% and 42\%, respectively. On the other hand, DJ's news also has keywords like GM. In this paper, we use the ALERTs, the HEADLINEs, and the titles of DJ's news.

\begin{figure}[!t]
\centering
\includegraphics[width=2.5in]{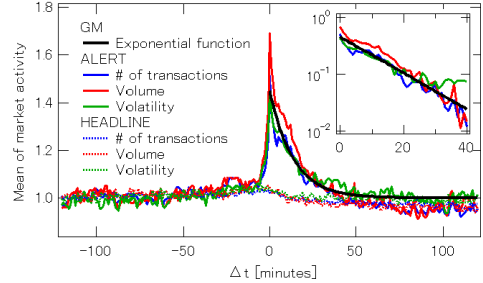}
\caption{Market activities of GM stock around GM's ALERT and HEADLINE are displayed. When $\Delta t=0$, news articles with keyword ``GM.N'' are displayed on RTRS's electronic trading platform. Upper right side figure shows semilogarithmic graph. Solid and dashed lines show market response to ALERT and HEADLINE, respectively. Black lines follow exponential function ($=0.45 \exp(-0.073 \Delta t)+1$).}
\label{fig_1}
\end{figure}

\begin{figure*}[!t]
\centering
\includegraphics[width=5in]{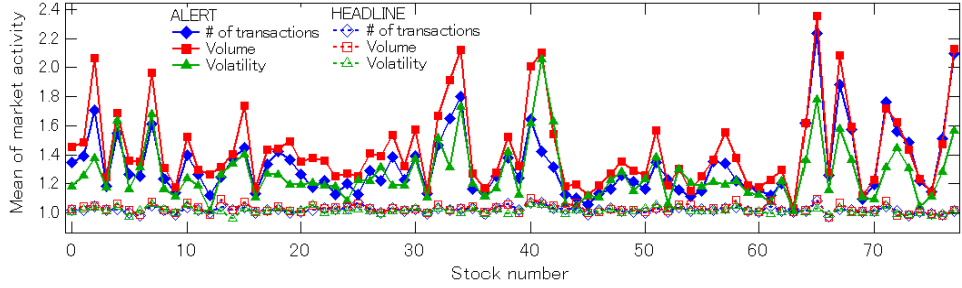}
\caption{Market activities of 78 stocks for three minutes after ALERT and HEADLINE were displayed. Ticker of each stock number is in Table~\ref{tabl_1}.}
\label{fig_2}
\end{figure*}

\begin{table*}[!ht]
\caption{Ticker that corresponds to stock number in Fig.~\ref{fig_2}.}
\begin{tabular}{llllllllllll}
\hline
Number	&Ticker &Number &Ticker &Number &Ticker &Number &Ticker &Number &Ticker &Number &Ticker\\
\hline
NYSE	&	& 	&	&	&	&	&	&	&	&	&\\
\hline
0	&AA	&1	&ABT	&2	&AET	&3	&AIG	&4	&AMD	&5	&APA\\
6	&APC	&7	&AXP	&8	&BA	&9	&BAC	&10	&BBT	&11	&BBY\\
12	&BHI	&13	&BK	&14	&BLK	&15	&BMY	&16	&C	&17	&CAT\\
18	&CHK	&19	&COF	&20	&COP	&21	&CVS	&22	&DD	&23	&DIS\\
24	&DOW	&25	&EMC	&26	&F	&27	&FCX	&28	&GD	&29	&GE\\
30	&GM	&31	&GS	&32	&HAL	&33	&IBM	&34	&JCP	&35	&JNJ\\
36	&JPM	&37	&KO	&38	&LLY	&39	&LMT	&40	&MDT	&41	&MO\\
42	&MRK	&43	&MRO	&44	&NEM	&45	&NOC	&46	&NOK	&47	&PEP\\
48	&PFE	&49	&PG	&50	&PNC	&51	&S	&52	&SLB	&53	&TWX\\
54	&UNH	&55	&UPS	&56	&USB	&57	&UTX	&58	&VIP	&59	&VLO\\
60	&WFC	&61	&WLP	&62	&WMT	&63	&XOM 	& 	&	&	&\\
\hline
NASDAQ	&	& 	&	&	&	&	&	&	&	&	&\\
\hline
64	&AAPL	&65	&AMGN	&66	&AMZN	&67	&BIIB	&68	&BRCM	&69	&CMCSA\\
70	&CSCO	&71	&DELL	&72	&DISH	&73	&EBAY	&74	&INTC	&75	&MSFT\\
76	&QCOM	&77	&YHOO	&	&	&	&	&	&	&	&\\
\hline
\end{tabular}
\label{tabl_1}
\end{table*}

\section{Market reaction to ALERT and HEADLINE articles}
To observe intraday market reaction to news, we measure market activities by volatility, the number of transactions, and transaction volume every minute for each stock. Volatility is defined by the absolute value of stock price log-return for one minute:

\begin{equation}
V'(d,t)=|\log{P(d,t+1)}-\log{P(d,t)}|,
\end{equation}

\noindent
where $d$ and $t$ express the date and the time of day [minutes] (e.g., $d=$5/18/2015, $t=$9:30 a.m.), respectively. 

Market activities have seasonal and daytime variations. We remove them from typical market cycles to correctly estimate the market impact on market activities for a day by introducing the normalized volatility, the normalized number of transactions, and the normalized volume as follows:

\begin{eqnarray}
V(d,t)=\frac{\frac{V'(d,t)}{\langle V'(d,t) \rangle_{d}}}{\langle \frac{V'(d,t)}{\langle V'(d,t) \rangle_{d}} \rangle_{t}}, \\
N(d,t)=\frac{\frac{N'(d,t)}{\langle N'(d,t) \rangle_{d}}}{\langle \frac{N'(d,t)}{\langle N'(d,t) \rangle_{d}} \rangle_{t}}, \\
Vol(d,t)=\frac{\frac{Vol'(d,t)}{\langle Vol'(d,t) \rangle_{d}}}{\langle \frac{Vol'(d,t)}{\langle Vol'(d,t) \rangle_{d}} \rangle_{t}},
\end{eqnarray}

\noindent
where $N'(d,t)$ and $Vol'(d,t)$ are the number of transactions and their volume at time $t$ on date $d$. Since $\langle \cdots \rangle_{d}$ expresses the mean on date $d$, daily seasonality is removed from the market activities by the first term in the equations. $\langle \cdots \rangle_{t}$ also expresses the mean at time $t$ in all the sample periods. The second term removes the intraday cycles of the market activities.

Next, we investigate the intraday market reaction to news that was displayed on RTRS's electronic trading platform. We observe three different market activities of GM stock in NYSE at time $\Delta t$ (i.e., $V(\Delta t)$, $N(\Delta t)$, $Vol(\Delta t)$), knowing that there was an ALERT or a HEADLINE with ``GM.N'' at time $\Delta t=0$. Fig.~\ref{fig_1} shows the mean of the market activities: $\langle V(\Delta t) \rangle$, $\langle N(\Delta t) \rangle$, $\langle Vol(\Delta t) \rangle$. In the ALERT case, the mean jumped about 60\% at time $\Delta t=0$ and slowly decayed in an exponential function ($=0.45 \exp(-0.073 \Delta t)+1$). On the other hand, the mean hardly moved when a HEADLINE was displayed.

We also investigate the intraday market reaction to the news of 64 NYSE stocks and 14 NASDAQ stocks in Table~\ref{tabl_1}. For each stock, the numbers of ALERTs and HEADLINEs are over 500 articles, and their total exceeds 3000 articles for the entire sample period. Fig.~\ref{fig_2} shows the conditional mean of the market activities of each stock for three minutes after news was displayed: $\langle V(\Delta t)|0 \le \Delta t < 3 \rangle$, $\langle N(\Delta t)|0 \le \Delta t < 3 \rangle$, $\langle Vol(\Delta t)|0 \le \Delta t < 3 \rangle$. In the ALERT case, we observe a jump in market activities in almost all the stocks. The mean of these jumps is 36.5\%. On the other hand, none of the stocks responded greatly to HEADLINE. These results suggest that we need to distinguish news that attract a lot of attention from market participants and news that receive little attention, and focus on news attracting a lot of attention in assessing the market response to such news. For the following sections, we examine the statistical laws regarding linguistic similarity among news articles, and propose measures for the novelty of a news article and for the topicality of an article.

\section{Similarity among news articles}
We use Inverse Document Frequency (IDF) and cosine similarity to measure the similarity among news articles. Such stop-words as ``and,'' ``with,'' and ``the'' are not good keywords to measure similarity, unlike such less common words as ``Chevrolet,'' ``antitrust,'' and ``bankrupt.'' IDF, which is a popular measure to determine whether a term is common or rare across all the articles in Natural Language Processing, is defined as a logarithm of the ratio of the total number of articles in a news dataset to the number of articles containing the given word in this paper.

Let $A=\{a_1,\cdots ,a_n\}$ be a set of articles and $W=\{w_1,\cdots ,w_m\}$ be a set of distinct words occurring in $A$. An article is represented as $m$-dimensional vectors $\vec{w_a}$. As mentioned previously, we use the $idf$ value as word weights and describe the vectors as follows:

\begin{eqnarray}
\vec{w_a}=(\delta (a,w_1)idf(w_1), \cdots , \delta (a,w_m)idf(w_m)),  \nonumber \\
\Biggl\{
\begin{array}{ll}
\delta (a,w_k)=1 & (w_k\in a) \\
\delta (a,w_k)=0 & (w_k\notin a) 
\end{array}
.
\end{eqnarray}

When articles are represented as vectors, the similarity of two articles corresponds to the correlation between the vectors. This is quantified as the cosine of the angle between vectors: the so-called cosine similarity. Given two articles, $a_i$ and $a_j$, their cosine similarity is

\begin{equation}
SIM(a_i,a_j)=\frac{\vec{w_{a_i}}\vec{w_{a_j}}}{|\vec{w_{a_i}}||\vec{w_{a_j}}|}.
\end{equation}

\noindent
As a result, the cosine similarity is bounded between $[0,1]$.

\begin{figure}[!t]
\centering
\includegraphics[width=2.2in]{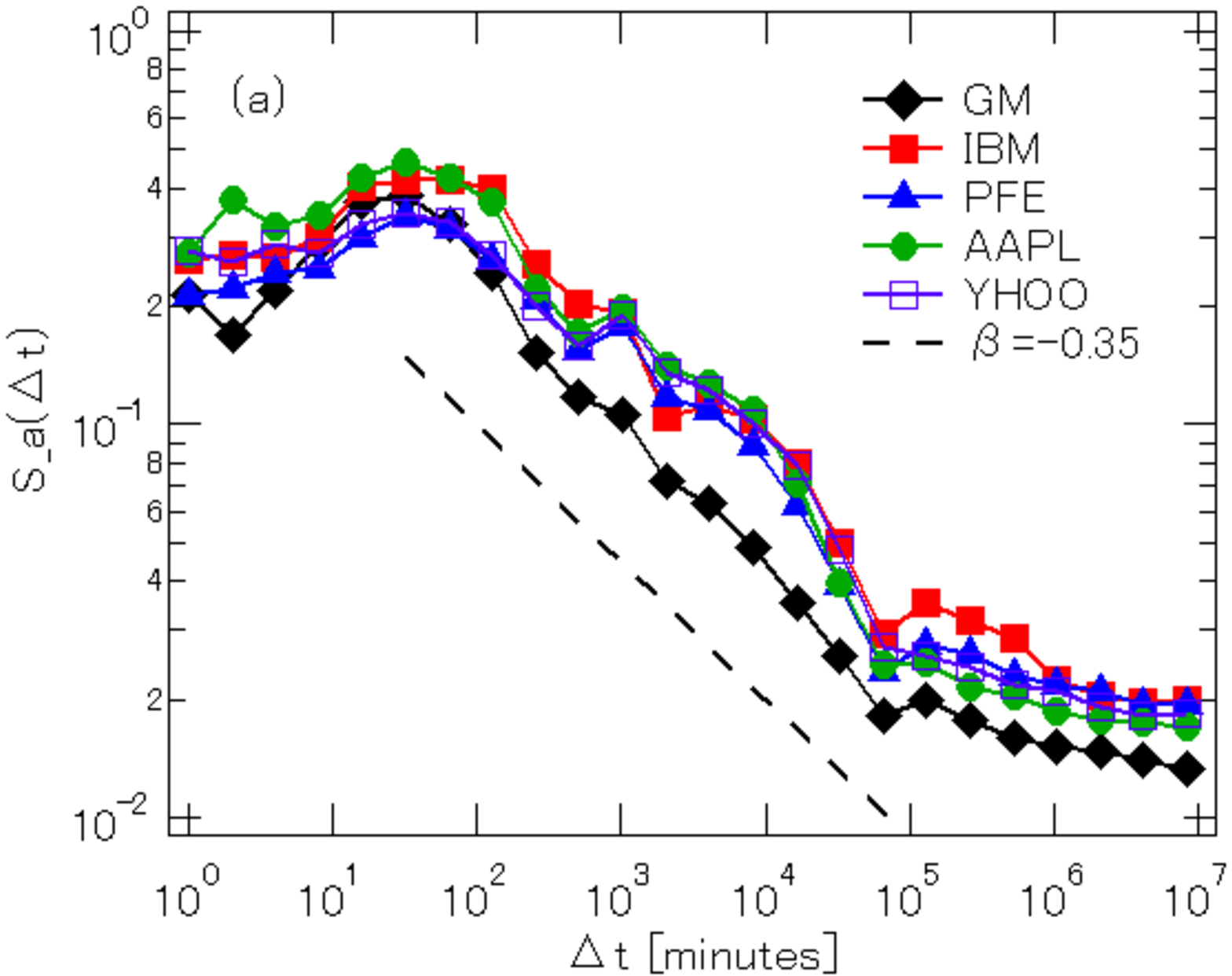}
\includegraphics[width=2.2in]{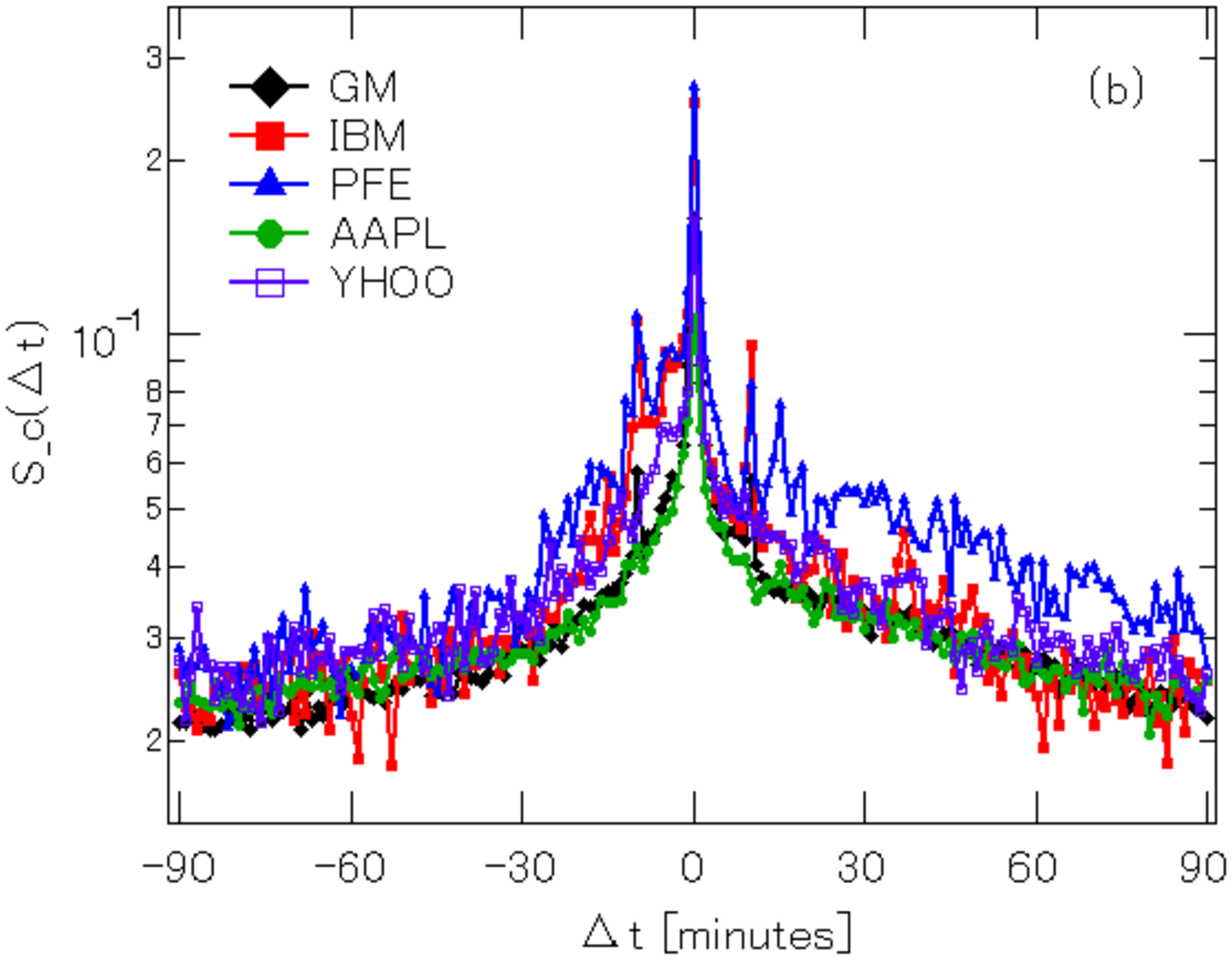}
\caption{(a) Auto cosine similarity function, (b) Cross cosine similarity function of news articles for GM, IBM, PFE, AAPL, and YHOO. Dashed line follows a power law with exponent $-0.35$.}
\label{fig_3}
\end{figure}

We also investigate the similarity among news articles in a time direction. Function $S_a (\Delta t)$ expresses the mean of the cosine similarity between the articles at different times $t$ and $t+\Delta t$. Throughout this paper, we call $S_a (\Delta t)$ the auto cosine similarity function for convenience. Fig.~\ref{fig_3}(a) shows the auto cosine similarity functions of RTRS news articles with keywords ``GM.N,'' ``IBM.N,'' ``PFE.N,'' ``AAPL.O,'' and ``YHOO.O'' (PFE.N and AAPL.O stand for Pfizer Inc. and Apple Inc., where .O means NASDAQ). The functions are almost constant, $S_a (\Delta t)=0.3$, at $\Delta t \le 200$ minutes. After that, the functions decay slowly until about $\Delta t=5 \times 10^5$ minutes $\approx 1$ year and are $S_a (\Delta t) \approx 0.02$ at $\Delta t=1$ year. The decay follows a power law, $S_a (\Delta t) \propto \Delta t^{-0.35}$ when $10^2 \le \Delta t \le 10^5$ minutes. These results suggest that news content tends to be remembered for several months.

We next focus on the similarity between news articles in the cross-sectional direction. Function $S_c (\Delta t)$, which expresses the mean of cosine similarity between RTRS's news at time $t$ and the news of other news agencies at time $t+\Delta t$, is called the cross cosine similarity function throughout this paper for convenience. Fig.~\ref{fig_3}(b) shows the cross cosine similarity functions for the articles with keywords ``GM.N,'' ``IBM.N,'' ``PFE.N,'' ``AAPL.O,'' and ``YHOO.O.'' The functions decay sharply compared to the auto cosine similarity function, and $S_c (\Delta t) \le 0.03$ at $|\Delta t| \ge 60$ minutes. A similarity peak is observed around $\Delta t=0$, $S_c (\Delta t) \approx 0.3$. This value is almost the same as the auto cosine similarity function at $\Delta t \le 200$ minutes, suggesting that multiple news agencies tend to simultaneously report similar news.

\section{Novelty and topicality detection}
Investors seek to forecast what will happen in the near future, and buy and sell securities based on such forecast. Therefore, it is important to distinguish between anticipated and unanticipated news. In this section, we first introduce the novelty measure for a news article and check whether novel news article identified by this measure using initial and follow-up news articles. On the other hand, even if a particular piece of news is unanticipated, market responses differ depending on the importance of that piece of news to investors. We assume a news article attracts a lot of attention and thus is highly topical if it is simultaneously reported by multiple news agencies and read by many investors who acquire news from those agencies. Based on this assumption, we next create the topicality measure for a news article. We also check whether topical news articles are caught by this measure using ALERTs and HEADLINEs.

News articles about common topics frequently use common words. By applying this characteristic, we define the novelty of news article $a_t$ at time $t$ by counting the number of linguistically similar news articles reported before the article $a_t$ as follows:

\begin{equation}
Nov(a_t)= \sum_{0 < \Delta t \le \tau}SIM(a_t,a_{t-\Delta t}),
\end{equation}

\noindent
when news articles at time $t$ and $t-\Delta t$ exist in a news dataset. Novelty is high when $Nov(a_t)$ is close to $0$. In this paper, we set maximum time lag $\tau$ to one week at which the auto cosine similarity function is around $0.1$ (Fig.~\ref{fig_3}(a)).

\begin{figure}[!t]
\centering
\includegraphics[width=2.5in]{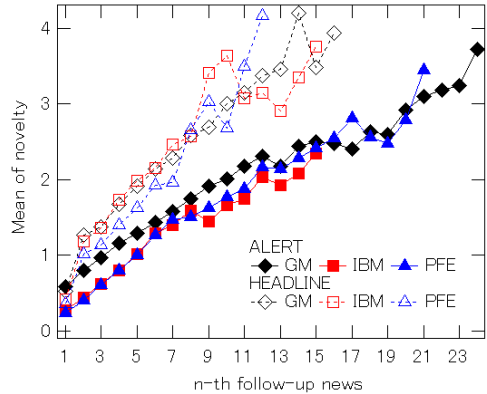}
\caption{Mean of novelty of ALERT and HEADLINE follow-up news articles for GM, IBM, and PFE.}
\label{fig_4}
\end{figure}

We check whether novel news article identified by novelty $Nov(a_t)$ using RTRS's follow-up articles for GM, IBM, and PFE that are included in our news dataset. Fig.~\ref{fig_4} shows the mean of $Nov(a_t)$ that is conditioned by the number of follow-ups. This conditional mean increases in proportion to the number of follow-ups. In Fig.~\ref{fig_4}, we compared the conditional mean for the ALERT and HEADLINE follow-ups. The novelty of ALERT is higher than that of HEADLINE except for initial news. 

Next, by applying the results of the cross cosine similarity function in Fig.~\ref{fig_3}(b), we define the topicality of news article $a_{t,k}$ at time $t$ at given news agency $k$ by counting the number of news articles which have similar content to the original news $a_{t,k}$ but are delivered by other news agencies as follows:

\begin{equation}
Top(a_{t,k})= \sum_{j\neq k, j\in K}SIM(a_{t,k},a_{t,j}),
\end{equation}

\noindent
when news articles $a_{t,k}$ and $a_{t,j}$ exist in a news dataset, where $K=\{k_1, \cdots, k_l\}$ is a set of news agencies. Topicality is high when $Top(a_{t,k})$ is large. Since topical news is actually reported by multiple news agencies at almost the same time, we consider the 30-minute periods before and after time $t$ as equal to time $t$. The cross cosine similarity function at 30 minutes is smaller than $0.05$, as shown in Fig.~\ref{fig_3}(b).

We check whether topical news articles are caught by topicality $Top(a_{t,k})$ comparing ALERT with HEADLINE. Table~\ref{tabl_2} expresses the means of topicality $Top(a_{t,k})$ of RTRS's ALERT and HEADLINE for GM, IBM, and PFE. The topicality of ALERT exceeds that of HEADLINE.

\begin{table}[!ht]
\caption{Mean of topicality $Top(a_{t,k})$ of RTRS's news for GM, IBM, PFE, AAPL, and YHOO.}
\begin{tabular}{llllll}
\hline
&GM	&IBM	&PFE	&AAPL	&YHOO\\
\hline
ALERT		&0.696	&0.863	&0.969	&0.778	&0.778\\
HEADLINE	&0.175	&0.171	&0.245	&0.106	&0.119\\
\hline
\end{tabular}
\label{tabl_2}
\end{table}

\section{Difference in market reaction by novelty and topicality of news}
\begin{figure}[!t]
\centering
\includegraphics[width=2.5in]{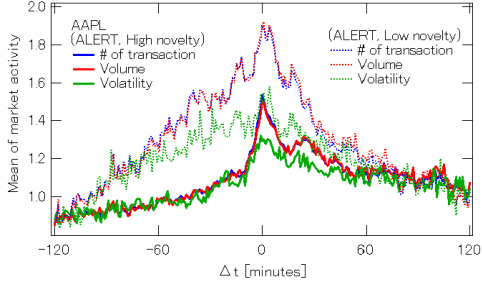}
\caption{Market activities of AAPL stock around high and low novelty ALERTs with ``AAPL.O''. Solid lines show market activities when novelty of ALERT $Nov(a_t)$ is above average. Dashed lines show market activities when novelty of ALERT $Nov(a_t)$ is below average.}
\label{fig_5}
\end{figure}

Using novelty $Nov(a_t)$ and topicality $Top(a_{t,k})$ of news, we investigate the intraday market reactions to both novel and topical news. Fig.~\ref{fig_5} shows the market activities (i.e., volatility $\langle V(\Delta t) \rangle$, number of transactions $\langle N(\Delta t) \rangle$, and transaction volume $\langle Vol(\Delta t) \rangle$ defined by Eqs.~(2)-(4)) of AAPL stock before and after ALERT with ``AAPL.O'' was reported. When $Nov(a_t) \ge \langle Nov \rangle$, market activities sharply increased just after the ALERT was reported at time lag $\Delta t=0$. When $Nov(a_t) < \langle Nov \rangle$, the market has already responded to the previous ALERTs and HEADLINEs before additional current ALERT occurs.

\begin{figure}[!t]
\centering
\includegraphics[width=2.5in]{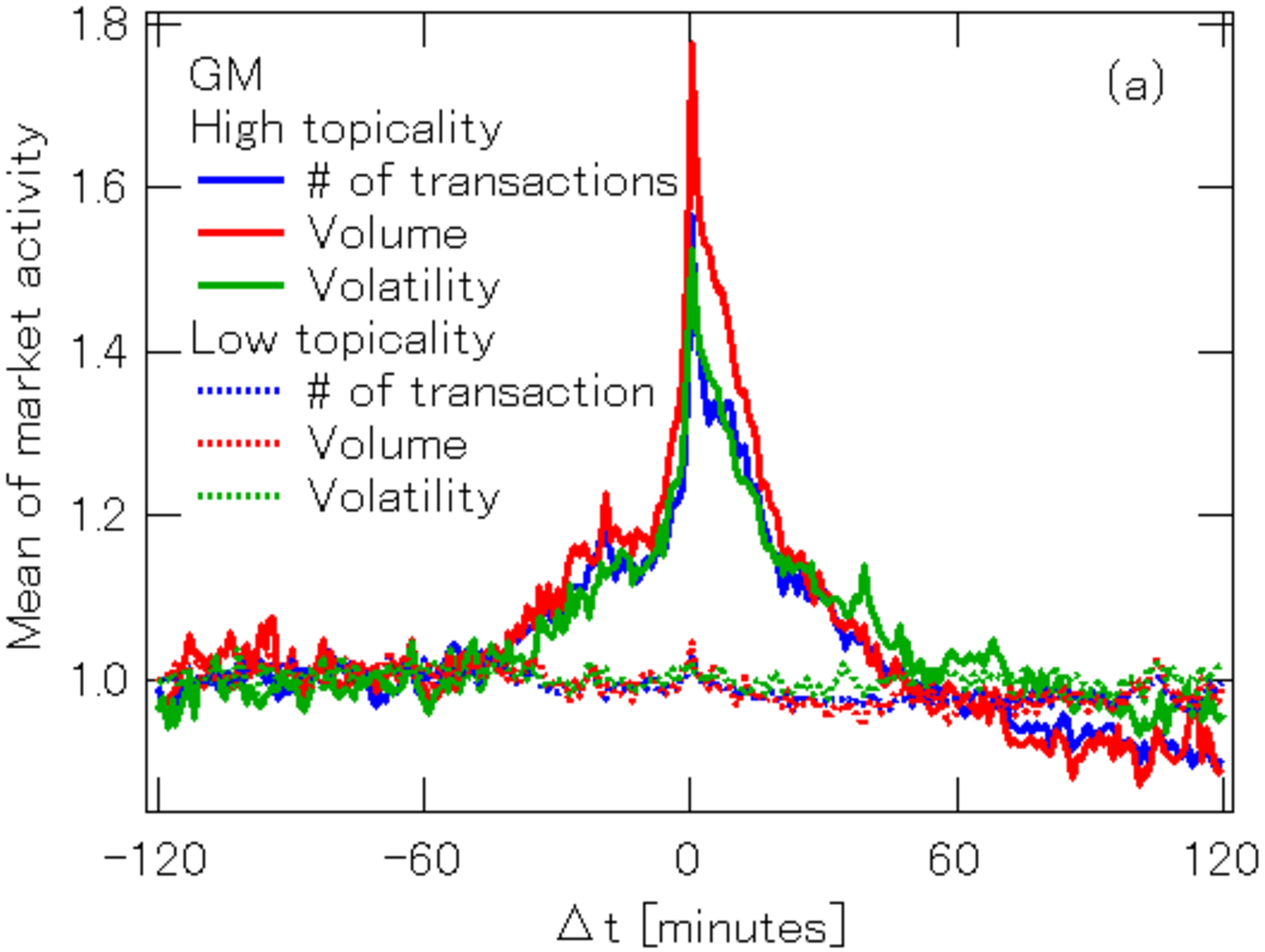}
\includegraphics[width=2.5in]{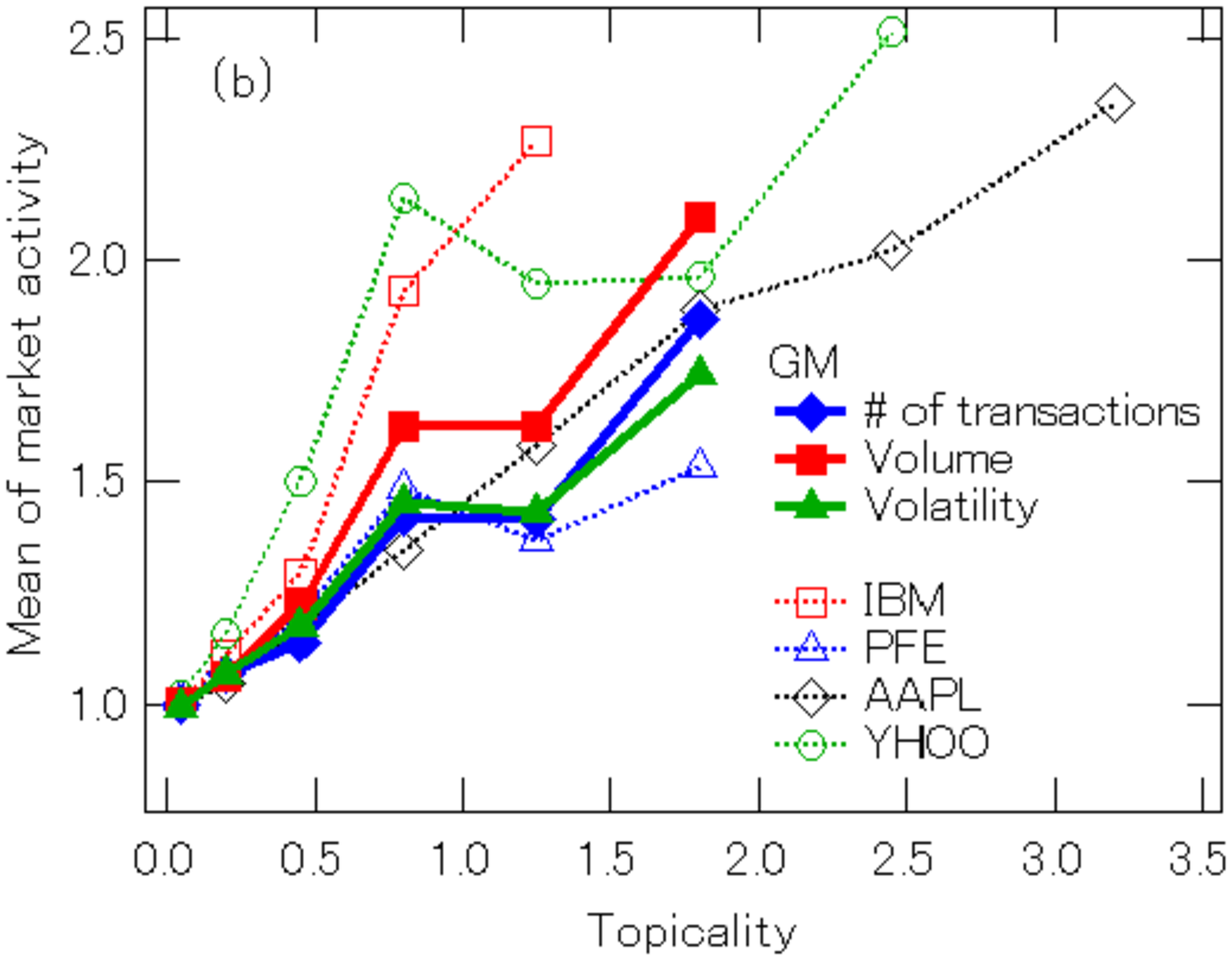}
\caption{(a) Market activities of GM stock around high and low topicality news with ``GM.N''. Solid lines show market activities when topicality of news $Top(a_{t,k})$ is above average. Dashed lines show market activities when topicality of news $Top(a_{t,k})$ is below average. (b) Relationship between topicality and mean of market activity for three minutes just after news was reported for GM, IBM, PFE, AAPL, YHOO. For IBM, PFE, AAPL, YHOO, market activity expresses mean of number of transaction, volume, and volatility.}
\label{fig_6}
\end{figure}

We investigate the relationship between market reaction and topicality $Top(a_{t,k})$ of news (i.e., ALERT and HEADLINE). As shown in Fig.~\ref{fig_6}, when $Top(a_{t,k}) \ge \langle Top \rangle$, the market responds greatly; when $Top(a_{t,k}) < \langle Top \rangle$, the market tends to avoid responding to the news article. The size of the response three minutes just after the news article was reported (i.e., $\langle V(\Delta t)|0 \le \Delta t < 3 \rangle$, $\langle N(\Delta t)|0 \le \Delta t < 3 \rangle$, $\langle Vol(\Delta t)|0 \le \Delta t < 3 \rangle$) is proportional to its news topicality $Top(a_{t,k})$.

\section{Conclusion}
We observed that the stock market strongly responds to the ALERTs that were displayed on RTRS's electronic trading platform. On the other hand, none of the stocks greatly responded to the HEADLINEs through which most news articles are reported. These results suggest that we need to measure the importance of news to predict market responses to it. In this paper, we focused on an indicator to measure the degree to which a particular news article is novel, as well as an indicator to measure the degree to which a particular news article acquires attention from investors. The novelty measure is obtained by comparing a news article with other news articles reported before that article in terms of linguistic similarity. On the other hand, we say a news article attracts a lot of attention and thus is highly topical if it is simultaneously reported by other news agencies and read by many investors who acquire news from those agencies. The topicality measure for a news article is obtained by counting the number of news articles which have similar content to the original news article but are delivered by other news agencies. In order to check whether novel or topical news articles are caught by these indicators, we observed that the novelty of follow-up news is lower than that of initial news and confirmed that the topicality of ALERT exceeds HEADLINE.

We found the characteristics of intraday market reactions to both novel and topical news. For a news article with high novelty, market activities (i.e., number of transactions, volume, volatility) sharply increased just after the news article was reported. On the other hand, for a news article with low novelty, market activities have already increased based on past similar news before the news article was reported. The increase of market activities based on news is proportional to its topicality.

By these results, we can empirically relate price movements to particular news to find convincing supportive evidence for efficient market hypothesis. Exogenous shocks often trigger or burst financial bubbles. Future work will investigate the characteristics of novel and topical news that cause bubbles to burst.

\section*{Acknowledgment}
This work was partially supported by JSPS KAKENHI Grant Number 24710156.



%

\end{document}